\documentclass[aps,pra,twocolumn,showpacs,amsmath,amssymb,floatfix]{revtex4}
\usepackage{graphicx} %Include figure files
\usepackage{dcolumn} %Align table columns on decimal point
\usepackage{bm} %bold math
\usepackage{epsfig}
\usepackage{amsmath}
\usepackage{amssymb}
\usepackage{float}
\usepackage{url}
\usepackage{color}
\usepackage{ulem}
\begin{document}
\newcommand{\bsy}[1]{\mbox{${\boldsymbol #1}$}} %bold greek
\newcommand{\bvecsy}[1]{\mbox{$\vec{\boldsymbol #1}$}} %bold greek with vectors
\newcommand{\bvec}[1]{\mbox{$\vec{\mathbf #1}$}} %bold vectors (no greeks)
\newcommand{\btensorsy}[1]{\mbox{$\tensor{\boldsymbol #1}$}} %bold greek tensor
\newcommand{\btensor}[1]{\mbox{$\tensor{\mathbf #1}$}} %bold tensor (no greeks)
\newcommand{\tensorId}{\mbox{$\tensor{\mathbb{\mathbf I}}$}} %unitary tensor
\newcommand{\be}{\begin{equation}}
\newcommand{\ee}{\end{equation}}
\newcommand{\bea}{\begin{eqnarray}}
\newcommand{\eea}{\end{eqnarray}}
\newcommand{\e}{\mathrm{e}}
\newcommand{\arccot}{\mathrm{arccot}}
\newcommand{\arctanh}{\mathrm{arctanh}}

\title{Negative refraction and rotons in the relativistic Bose gas}
 
\author{D. M. Reis$^{1}$, S. B. Cavalcanti$^{2}$, and C. A. A. de Carvalho$^{1,3}$}

\affiliation{
$^1$Centro Brasileiro de Pesquisas F\'{\i}sicas - CBPF, Rua Dr. Xavier Sigaud 150, Rio de Janeiro - RJ, 22290-180, Brazil\\
$^2$Instituto de F\'{\i}sica, Universidade Federal de Alagoas - UFAL, Cidade Universit\'aria, Macei\'o - AL, 57072-970, Brazil\\
$^3$Instituto de F\'{\i}sica, Universidade Federal do Rio de Janeiro - UFRJ, Caixa Postal  68528, Rio de Janeiro - RJ, 21945-972, Brazil}

\begin{abstract}
We investigate the dispersion of a classical electromagnetic field in a relativistic ideal gas of charged bosons using scalar quantum electrodynamics at finite temperature and charge density. We derive the effective electromagnetic responses and the electromagnetic propagation modes that characterize the gas as a left-handed material with negative effective index of refraction $n_{\rm eff}=-1$ below the transverse plasmon frequency. In the condensed phase, we show that the longitudinal plasmon dispersion relation exhibits a roton-type local minimum that disappears at the transition temperature.

\end{abstract}


\date{\today}

\maketitle

The relativistic Bose gas is an ideal gas of charged bosons and antibosons whose dispersion relation is $E_{\pm}(\vec{p})= \pm \sqrt{\vec{p}^2 c^2+ m^2c^4}$ ($+$ bosons, $-$ antibosons). A chemical potential $ - m c^2 \le \xi \le +m c^2$ 
is used to fix the conserved charge, proportional to the number of bosons minus antibosons. The system undergoes a phase transition, forming a Bose-Einstein condensate below a critical temperature $T_c$ \cite{lieb}.

In this letter, we address the interaction of a classical electromagnetic field with the relativistic Bose gas, using a semiclassical version of scalar quantum electrodynamics at finite temperature and charge density. We derive the effective electric permittivity, the effective magnetic permeability, and the electromagnetic modes of propagation of the gas, both in the normal and condensed phases.

As a result: i) we establish the gas as a left-handed material below the transverse plasmon frequency; ii) we show that it supports longitudinal and transverse plasmons, and a photonic mode that propagates without loss with the velocity of light in vacuum, which we use to characterize it as a medium with negative effective index of refraction $n_{\rm eff}=-1$ below the transverse plasmon frequency; iii) we check for signatures of the condensed phase in the dispersion relations of the electromagnetic propagation modes.

Left-handed materials are media whose effective permittivity and effective permeability in Maxwell's equations are simultaneously negative within some frequency range, where the phase velocity lies opposite to the Poynting vector. In a series of articles \cite{AragaoPRDS2016,ReyezEPL2016,JPP,ReisAdP2018,AragaoJOSA2020}, the relativistic electron gas was established a {\it natural} left-handed material, in contrast to {\sl artificially} engineered metamaterials \cite{Veselago,Pendry,engheta,handbook}. The key ingredient for that was the relativistic nature of the model, which amplified magnetic responses, setting them on equal standing as electric ones. That was one of the motivations to study the relativistic Bose gas, and search for similar behavior.

The other motivation was the possibility of searching for structure in the condensed phase of the gas, inspired by the physics of superfluids such as liquid $^{4}{\rm He}$, described by self-interacting charged scalars. There, the observation via neutron scattering \cite{clarendon} of collective phonon-roton modes in the (condensed) superfluid phase was a major discovery. Such collective excitations were ultimately responsible for superfluidity, according to the seminal work of Landau \cite{landau}. 

We have found structures similar to superfluid rotons in the dispersion relation of the longitudinal plasmon mode of the relativistic Bose gas, which exhibits a roton-type local minimum that disappears at the critical temperature, whose gap energy vanishes at $T_c$. This strongly suggests that we are seeing rotons in the charge density oscillations that disorder the system, and drive it into the normal phase.

The action for scalar quantum electrodynamics at finite temperature and charge density is \cite{Kapusta}
\be
S= \int_0^\beta \!\! d\tau\!\! \int\!\! d^3 x \, (\frac{1}{4}F_{\mu\nu}F_{\mu\nu}+\bar{D}_\mu\phi^* \bar{D}_\mu\phi+m^2\phi^*\phi),
\ee
where $k_B=\hslash=c=1$, and $\beta=1/T$. $A_\mu$ is an electromagnetic field, $\phi$ is a complex scalar describing spin-0 charged bosons of mass $m$ and charge $e$, $F_{\mu\nu}=\partial_\mu A_\nu - \partial_\nu A_\mu$, and $\bar{D}_\mu\phi=(\bar{\partial}_\mu-ie A_\mu)\phi$, with $\bar{\partial}_\mu\equiv(\partial_4-\xi, \partial_i)$

The effective action for a classical $A_\mu$ that incorporates the electromagnetic response of the gas may be obtained from the free energy, i.e., minus the logarithm of an Euclidean functional integral over scalars satisfying $\phi(0,\vec{x})=\phi(\beta,\vec{x})$ 
\be
\label{eq: S bosons}
S_{\rm eff}=\frac{1}{4}F_{\mu\nu}F_{\mu\nu}+\text{Tr}\ \ln\left(-\bar{D}^2+m^2\right).
\ee
The extremum condition leads to Maxwell's equation $\partial_\mu F_{\mu\nu}=J_\nu(x)=
\delta {\rm Tr}\ln (-\bar{D}^2+m^2)/\delta A_\nu (x)$. The current density has vacuum and medium contributions, $J_\nu=J_\nu^{(v)}+J_\nu^{\rm (m)}$. The latter defines $P_{\mu\nu}$, through $iq_\mu{P}_{\mu\nu}(q)=-J^{\rm (m)}_\nu (q)$, whose components are the polarization $\vec{P}, P_{4j}=iP^j$, and the magnetization $\vec{M}, P_{ij}=-\epsilon_{ijk}M^k$. 

Expanding for weak classical fields $A_\mu$ up to quadratic terms yields a linear response approximation ${J}^{\rm (m)}_{\mu}={\Pi}_{\mu\nu}(q){A}_\nu(q)$, where ${\Pi}_{\mu\nu}$ is the field theory polarization tensor defined below. One may show \cite{AragaoPRDS2016} that $q^2{P}_{\mu\nu} = \Pi_{\mu\sigma} F_ {\nu\sigma}- \Pi_{\nu\sigma} F_{\mu\sigma}$, and define ${H}_{\mu\nu}\equiv {F}_{\mu\nu}+{P}_{\mu\nu}$, whose components are the electric displacement $\vec D$, ${H}_{4j}= iD^j$, and the magnetic induction $\vec H$, ${H}_{ij}= - \epsilon_{ijk} H^k$. 

The field theory polarization tensor ${\Pi}_{\mu\nu}$ contributes two terms to the induced current, the tadpole and the thermal bubble Feynman graphs 
\bea
\label{eq: Tensor Pol}
&&{\Pi}_{\mu\nu}(q) =-\frac{2e^2}{\beta}\sum_{n=-\infty}^{+\infty}\int \frac{d^3p}{(2\pi)^3}\frac{\delta_{\mu\nu}}{\bar{p}^2+m^2}\nonumber\\
&&+\frac{e^2}{\beta}\sum_{n=-\infty}^{+\infty}\int \frac{d^3p}{(2\pi)^3}\frac{(2\bar{p}_\mu+q_\mu)(2\bar{p}_\nu+q_\nu)}{(\bar{p}^2+m^2)[(\bar{p}+q)^2+m^2]}.
\eea
The sum is over bosonic Matsubara frequencies $p_4=2n\pi T$. Writing ${\Pi}_{\mu\nu}={\Pi}_{\mu\nu}^{(v)}+{\Pi}_{\mu\nu}^{(m)}$, symmetry and gauge invariance \cite{IZ,AP} lead to ${\Pi}_{44} ^{(m)} = - q^2 {\cal B}$, ${{\Pi}_{4i} ^{(m)}} = - q^2 \hat{q}_4 \hat{q}_i {\cal B}$, ${{\Pi}_{ij} ^{(m)}} = - q^2\left[ \left (\delta_{ij} - \hat{q}_i\hat{q}_j\right ) {\cal{A}} + \delta_{ij} \hat{q}_4^2 {\cal{B}} \right ]$, and ${ \Pi}_{\mu \nu}^{(v)}=-(q^2\delta_{\mu \nu}-q_\mu q_\nu) \mathcal{C}(q^2)$, where $\hat{q}_{\mu} \equiv q_{\mu}/|\vec{q}|$, and $q^2=\vec q^2 + q_4^2$. The scalar functions $\mathcal{A} (|\vec q|, q_4)$, $\mathcal{B} (|\vec q|, q_4)$, and $\mathcal{C}(q^2)$ are given by integrals containing the Bose-Einstein occupation number for bosons and antibosons $n(p)=n^+(p)+n^-(p)=\left\{{\rm e}^{\beta(\omega_p-\xi)}-1\right\}^{-1}+\left\{{\rm e}^{\beta(\omega_p+\xi)}-1\right\}^{-1}$ 
%\be
%n_B(p)=\frac{1}{{\rm e}^{\beta(\omega _p-\xi)}-1}+\frac{1}{{\rm e}^{\beta(\omega %_p+\xi)}-1}.
%\ee
%May be written as $\mathcal{A} = I_a + [1-(3q^2/2|\vec{q}|^2)] I_b$, and $\mathcal{B} %= I_b$
 %\bea
 %\label{Afunc}
 %&&I_a =  \frac{e^2}{2 q^{2}}\text{Re}\!\!
 %\int \!\!\frac{d^3 p}{(2 \pi)^3 } \frac{n_B (p)}{\omega_p} \frac{3q^2+4m^2+4p\cdot %q}
 %{q^2 + 2 p \cdot q},\nonumber\\
 %\label{Bfunc}
 %&&I_b =  \frac{e^2}{q^{2}} \text{Re}\!\! \int \!\! \frac{d^3 p}{(2 \pi)^3} \frac{n_B %(p)}{\omega_p} \frac{|\vec{ q}|^2+4\omega_p^2-2p_4q_4+2\vec{p}\cdot\vec{ q}}{q^2 + 2 %p \cdot q},\nonumber
 %\eea

The tensors $H_{\mu\nu}$ and $P_{\mu\nu}$ provide the constitutive equations,  ${ D}_j=\epsilon_{jk} { E}_k + \tau_{jk}  { B}_k$, $ { H}_j= \nu_{jk}  { B}_k + \tau_{jk}  { E}_k$, with $\nu\equiv \mu^{-1}$. We may write $\epsilon_{jk}= \epsilon \delta_{jk} + \epsilon' \hat{q}_j \hat{q}_k$, $\nu_{jk}= \nu \delta_{jk} + \nu' \hat{q}_j \hat{q}_k$, and $\tau_{jk}= \tau \epsilon_{jkl} \hat{q}_l$. The eigenvalues of $\epsilon_{jk}$ are $\epsilon + \epsilon'$ and $\epsilon$. The eigenvector of $\epsilon+ \epsilon'$ is along $ \hat{q}$, thus longitudinal, whereas the two eigenvectors with eigenvalues $\epsilon$ are transverse to $ \hat{q}$. The same occurs for $\nu_{jk}$, with eigenvalues $\nu + \nu '$ and $\nu$. $\tau_{jk}$ is clearly transverse. 

a) For $T>T_c$, the permittivities and inverse permeabilities are determined by the three scalar functions  ${\cal A}^\ast$, ${\cal B}^\ast$, and ${\cal C}^\ast$. The asterisk denotes continuation to Minkowski space $q_ 4 \rightarrow i\omega - 0^+$ of the Euclidean scalar functions ${\cal{A}} (|\vec q|, q_4)$, ${\cal B} (|\vec q|, q_4)$ and ${\cal{C}} (\vec{q}^2 + q_4^2)$. From now on, we shall use the Minkowski definition $q^2 = \omega^2 - |\vec{q}|^2$ ($q^2_E \rightarrow - q^2_M$). 

b) For $T<T_c$, a part $\eta^{(c)}(T)$ of the density of charge $n^+ - n^-$ condenses in the ground state ($\vec{p}=0$), as Bose-Einstein condensation sets in for $\xi=m$. To account for the part that condenses, the Bose-Einstein distribution $n(p)$ must be modified to
\be
\label{nB1}
n(p)\rightarrow (2\pi)^3\eta^{(c)}(T)\delta^{(3)}(\vec{p})+n^\prime (p),
\ee
with $n^\prime (p)=\left\{{\rm e}^{\beta(\omega_p-m)}-1\right\}^{-1}+\left\{{\rm e}^{\beta(\omega_p+m)}-1\right\}^{-1}$. The modification has a contribution due to the charge density in the ground state, and another due to the particles in excited states. The longitudinal responses, $\epsilon_L=\epsilon + \epsilon'$ and $\nu_L=\nu + \nu'$, become
\bea
&& \epsilon_L(\omega,|\vec{q}|)=1+\mathcal{C}^*-\omega^2_e\left(\frac{q^2-4m^2}{q^4-4m^2\omega^2}\right) -\frac{q^2}{|\vec{q}|^2} \mathcal{B}^*_T, \nonumber \\
&& \nu_L(\omega,|\vec{q}|)=1-\frac{2\omega^2_e}{q^2}+2
\mathcal{C}^*+2\mathcal{A}^*_T-2\frac{\omega^2}{|\vec{q}|^2}
\mathcal{B}^*_T.
\eea
$\omega^2_e=e^2\eta^{(c)}/m$ is the longitudinal electric plasmon frequency; $\mathcal{A}^*_T$ and $\mathcal{B}^*_T$ are the functions ${\cal A}^\ast$ and ${\cal B}^\ast$ with the modification $n(p)\rightarrow n^\prime (p)$. 

c) For $T=0$, all the charge condenses, so $n_c^\prime (p)\rightarrow 0$, and the scalar functions $\mathcal{A}^*_T$ and $\mathcal{B}^*_T$ vanish. Neglecting the vacuum contribution ($\mathcal{C}^*<<1$), one obtains  
\bea
\label{eq: epsilon_bose}
&&\epsilon_L(\omega,|\vec{q}|)=1-\omega^2_e\left(\frac{q^2-4m^2}{q^4-4m^2\omega^2}\right), \nonumber \\
\label{eq: nu_bose}
&& \nu_L(\omega,|\vec{q}|)=1-\frac{2\omega^2_e}{q^2},
\eea
for $q^2 < 4m^2$. For $q^2 > 4m^2$, imaginary parts appear ${\rm Im} \epsilon_L (\omega,|\vec{q}|)= [e^2/48 \pi][(1 - (4m^2)/(q^2)]^{3/2}$.
As $|\vec{q}|\rightarrow 0$, the longitudinal responses at $T=0$ are the same as those obtained in \cite{AragaoPRDS2016} for the relativistic electron gas, $\epsilon_L=1-\omega_e^2/\omega^2$ and $\nu_L=1-2\omega_e^2/\omega^2$.

We now turn our attention to the propagation of collective modes in the gas, and evaluate how the medium affects the photon propagator in the Bose gas. Just as in the case of the relativistic electron gas, we obtain the photon propagator ${\Gamma}_{\mu\nu}^{-1}$ in the form
\be
\label{eq: photon_propagator}
{\Gamma}_{\mu\nu}^{-1}=\frac{\mathcal{P}^{L}_{\mu\nu}}{-q^2\epsilon_L}+\frac{\mathcal{P}^{T}_{\mu\nu}}{-q^2(\nu_L+1)}+\frac{\lambda}{q^2}\frac{q_\mu q_\nu}{q^2},
\ee
where we have introduced the projectors $\mathcal{P}_{\mu\nu}\equiv \mathcal{P}_{\mu\nu}^L+\mathcal{P}_{\mu\nu}^T=\delta_{\mu\nu}-q_\mu q_\nu/q^2$, with $\mathcal{P}^T_{ij}=\delta_{ij}-\hat{q}_i\hat{q}_j$, and $\mathcal{P}^T_{00}=\mathcal{P}^T_{0i}=0$, and $\lambda$ a gauge parameter. The poles of the photon propagator correspond to collective excitations and yield their dispersion relations. 

In the longitudinal propagator, there is a pole whenever the longitudinal electric permittivity vanishes $\epsilon_L(\omega,\vec{q})=0$, leading to a longitudinal plasmon mode. For $\epsilon_L(\omega,\vec{q})$ nonzero, Maxwell's equations lead to transverse fields $\vec{q}\cdot\vec{E}=0$ that vanish when contracted with $\mathcal{P}_{\mu\nu}^L$, so that $q^2=0$ is not realized in this case. The transverse propagator has a pole whenever $\mu^{-1}_L(\omega,\vec{q}) = \nu_L(\omega,\vec{q})=-1$, which corresponds to collective oscillations of the current density; it has another pole whenever $q^2=\omega^2-|\vec{q}|^2=0$, which corresponds to a photonic mode propagating with the speed of light $c$ in vacuum. Similar modes have already appeared in the relativistic electron gas \cite{ReisAdP2018}.

\begin{figure}[ht!]
	\begin{center}
		\includegraphics[width=70mm]{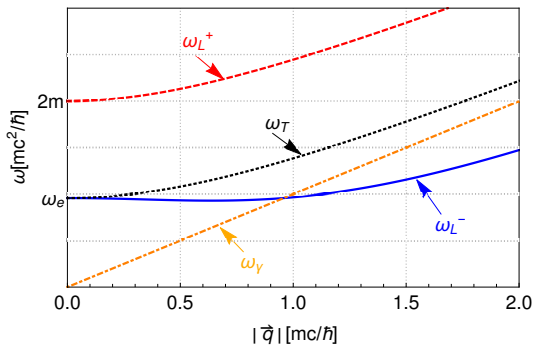}
\caption{\footnotesize Longitudinal $(\omega_{L})$, transverse  $(\omega_{T})$, and photonic $(\omega_\gamma)$ modes at $T=0$ and boson density $\eta=10 (mc/\hslash)^3$.}
		\label{fig3}
	\end{center}
\end{figure}

At $T=0$, the plasmon modes have longitudinal $\omega_L(|\vec{ q}|)$ and transverse $\omega_T(|\vec{q}|)$ dispersion relations, obtained from (\ref{eq: epsilon_bose}), given by 
\be
\label{eq: omega_L_boson}
\omega^2_{L\pm}=\frac{1}{2}(4m^2+\omega_e^2+2|\vec{q}|^2)\pm\frac{1}{2}\sqrt{(4m^2-\omega_e^2)^2+16m^2|\vec{q}|^2}, \nonumber 
\ee
\be
\label{eq: omega_T_boson}
\omega^2_T=\omega_e^2+|\vec{q}|^2.
\ee
The expression for $\omega^2_{L}$  has two branches: one beginning at $\omega_{L-}=\omega_e$, at $|\vec{q}|=0$; the other, which leads to pair creation, beginning at $\omega_{L+}=2m$, at $|\vec{q}|=0$. For $\omega_e^2<4m^2$, the longitudinal mode will show a local minimum at low values of $|\vec{q}|$, known as  {\sl negative dispersion} \cite{melrose}. This behavior will appear only in the relativistic regime, since in the non-relativistic limit our expression reduces to $\omega^2_L=\omega_e^2+|\vec{q}|^4/4m^2$, the same result obtained in \cite{hore1975} for a charged Bose gas within the random phase approximation. %Fig.\ref{fig1} shows the normalized longitudinal modes, $\omega_{L-}$,  at zero temperature, for some values of the electron density $\eta$.% 
Fig.\ref{fig3} shows the dispersion curves of the two possibilities for the longitudinal mode, and the transverse modes.

We stress that the responses discussed thus far are NOT the ones appearing in Maxwell’s equations. In order to find the exact analog of the situation proposed by Veselago \cite{Veselago}, we need the effective electric permittivity and effective magnetic permeability that appear in the two equations with sources. Following \cite{AragaoJOSA2020}, we may define the effective responses as 
\be
\mu_{\rm eff}=\frac{\mu|\vec{ q}|}{|\vec{ q}|-\omega\mu\tau};\,\,\,\,\,\,\,\,\, \epsilon_{\rm eff}=\epsilon+\frac{|\vec{ q}|}{\omega}\tau.
\ee
In terms of $\mathcal{A}^*$ and $\mathcal{B}^*$ ($\mathcal{C}^*<<1$), we have 
\be
\label{effresp0}
\epsilon_{\rm eff}= \nu_{\rm eff}= \mu_{\rm eff}^{-1} = \frac{\nu_L + 1}{2}= 1+\mathcal{A}^*-\frac{\omega^2}{|\vec{ q}|^2}\mathcal{B}^*.
\ee
Expression (\ref{effresp0}) immediately implies that $|n_{\rm eff}| = {\mu_{\rm eff}}\epsilon_{\rm eff} = 1$, for any temperature and chemical potential. It also implies $\epsilon_{\rm eff}= \nu_{\rm eff}<0$ as long as $\nu_L<-1$, which means that we will have left-handed behavior for frequencies below the transverse plasmon frequency $\omega_T (|\vec{q}|)$, ($\nu_L=-1$), for any temperature and chemical potential. In particular, for $T=0$, using (\ref{eq: nu_bose}), we obtain
\be
\label{effresp} 
\epsilon_{\rm eff}=\nu_{\rm eff}= \frac{\nu_L+1}{2} = \left(1- \frac{\omega_e^2}{q^2}\right), 
\ee
 which reduces, in the limit $|\vec{q}|\rightarrow 0$, to a Drude expression, $\epsilon_{\rm eff}=\nu_{\rm eff}= (1- \omega_e^2/\omega^2)$.
 
The existence of a photonic mode $\omega_\gamma = |\vec{q}|$ with the speed of light in vacuum is another indication that the modulus of the effective index of refraction of the gas is equal to one. Since, both $\epsilon_{\rm eff} < 0$ and $\nu_{\rm eff}< 0$, for $\omega_\gamma (|\vec{q}|)\leq \omega < \omega_T( |\vec{q}|)$, for any temperature and chemical potential, one may use Snell's law to show that $n_{\rm eff}=-1$ in that region, a negative refraction typical of a left-handed material \cite{Veselago,AragaoJOSA2020}. The shadowed sections of Fig.\ref{fig1} illustrate those regions for both $T=0$ and $T\ne 0$. Similar behavior was obtained in the analysis of the relativistic electron gas, confirming our hypothesis that the relativistic nature of the gas was the key ingredient to achieve left-handed behavior. 

{Fig.\ref{fig1}(a) shows that, in the condensed phase, the region where the gas exhibits left-handed behavior shrinks with increasing temperature, at least up to $T_c$. Somewhere above $T_c$  (see Fig.\ref{fig1}.(b)), the region expands and, for the case of ultra-relativistic densities $(\eta \ge (mc/\hslash)^3)$, it begins to expand at the critical temperature $T=T_c$, as illustrated in Fig.\ref{fig1}(c). Fig.\ref{fig2} shows the plasmon energy at $|\vec{q}| = 0$, normalized by its value at $T=0$, as a function of $T/T_c$ for various densities. 

In contrast to the non-relativistic case, the transverse plasmon frequency at $|\vec{q}| = 0$,  known as plasmon gap energy, is a function of temperature for relativistic densities. In fact, the two regimes just described correspond to $T < T_t$ and $T \ge T_t$, where $T_t$ is the temperature at which the transverse plasmon gap energy begins to increase linearly with temperature. This temperature increases with decreasing density, and has the critical temperature as its lower bound in the ultra-relativistic regime.
Numerical estimates for $\tau_t \equiv k_B T_t/m c^2, \tilde{\eta} \equiv \eta (\hbar/m c^2)$ give $(\tau_t,\tilde{\eta})=(5.47,10); (1.84,1); (0.74,0.1); (0.37,0.01)$. We note that similar behavior is observed in the relativistic electron gas, where $k_B T_t$ may be associated with the Fermi energy.

For ultra-relativistic densities, one may perform a high temperature hard thermal loop expansion \cite{Kapusta}, and derive analytic solutions for the transverse plasmon frequency $\omega_T(|\vec{q}|)$. At $|\vec{q}|=0$, we obtain $\omega^2_T(0)=\omega^2_e+e^2 T^2/9$, for $T\leq T_c$, and $\omega^2_T(0)=e^2 T^2/9$, for $T\ge T_c$. For $T<T_c$, using $\omega^2_e=e^2\eta^{(c)}/m$, and the ultra-relativistic expression $\eta^{(c)}=\eta(1-T^2/T_c^2)$, we obtain
\be
\label{omegaT}
\omega^2_T(0)=\frac{e^2\eta}{m}\left[1-\left(\frac{T}{T_c}\right)^2\right]+\frac{e^2T^2}{9},
\ee
which shows that the transverse plasmon frequency decreases with temperature in the condensed phase, and increases linearly in the normal phase, with $\omega_T(0)=e T/3$, for $T\ge T_c$. 
\begin{figure}[ht!]
	\begin{center}
		\includegraphics[width=85mm]{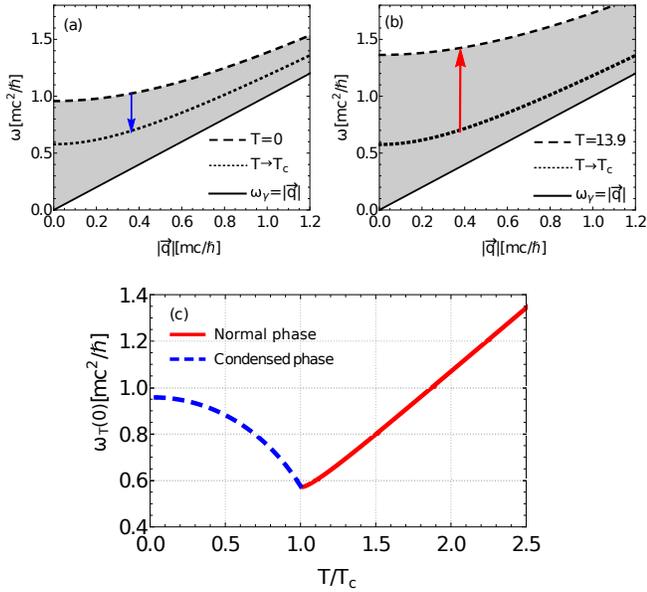}
\caption{\footnotesize  The shadowed regions illustrate where the RBG is a LHM: (a) in the condensed phase, the LHM region decreases with the temperature until we reach the critical temperature $T_c$; (b) above the critical temperature $T_c$, the LHM region starts to increase with the temperature; (c) transverse plasmon $\omega_T$ at $|\vec{q}|= 0$ as a function of the relative temperature. Calculations were performed for $\eta=10(mc/\hslash)^3$ and $T_c=5.47 mc^2/k_B$.}
		\label{fig1}
	\end{center}
\end{figure}

\begin{figure}[ht!]
	\begin{center}
		\includegraphics[width=70mm]{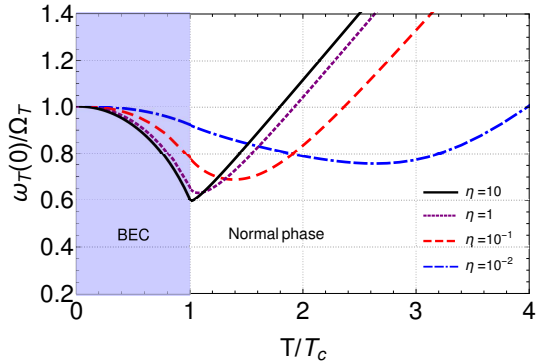}
\caption{\footnotesize  Normalized transverse plasmon frequency in the long-wavelength limit $(|\vec{q}|=0)$ as a function of the relative temperature for various densities [in units of $(mc/\hslash)^3$] of the Bose gas in units of $\Omega_T\equiv\omega_T(|\vec{q}|=0,T=0)$. In the shadowed region, below the critical temperature $T_c$, the gas is in the Bose-Einstein condensed phase (BEC). For each density $\eta$, the critical temperatures [in units of $(mc^2/k_B)$] are $T_c=5.47 (\eta=10)$, $T_c=1.72 (\eta=1)$, $T_c=0.53  (\eta=10^{-1})$, and $T_c=0.14 (\eta=10^{-2})$. }
		\label{fig2}
	\end{center}
\end{figure}
We now turn to the longitudinal plasmon mode. For $T>0$, we have to resort to numerical results. Fig.\ref{fig4}.(a) depicts the longitudinal dispersion relation $\text{Re}[\epsilon_L]=0$ for temperatures above and below the critical temperature $T_c$ of Bose-Einstein condensation. For $T>T_c$, the dispersion relation is similar to the case of the relativistic electron gas. For $T<T_c$, we observe a new kind of elementary excitation, with a local maximum and a local minimum analogous to the maxons and rotons in the spectrum of {\it neutral} superfluid $^{4}{\rm He}$ described by Landau \cite{landau}, who proposed the existence of two kinds of elementary excitations in a {\it neutral} superfluid: phonons, for low wavevectors, associated to acoustic waves; and rotons, gapped excitations at finite momentum $|\vec{q}|=|\vec{q}_{rot}|$, interpreted as vortices in the superfluid. 

The existence of roton-like structures has been predicted for condensates of dipolar particles \cite{santos,fischer,wilson,lahaye,bisset}, of nonpolar atoms under the action of an intense laser light \cite{odell}, and for Rydberg-excited condensates \cite{henkel}. Recently, these vortex-like quasi-particles have been observed for the first time in a Bose-Einstein condensate of ultra-cold atoms \cite{chomaz2018observation}. However, in a {\it charged} superfluid, it has been shown that the phonon mode of the neutral superfluid is pushed to a finite plasmon frequency $\omega_p$, whereas the roton mode is more or less unaffected \cite{hirashima1985,zhang1995}. In the charged case, the spectrum of the superfluid field shows a plasmon excitation that turns into a roton excitation, with a gap energy $\Delta(|\vec{q}_{rot}|)$ for higher $|\vec{q}|$. 
%\begin{figure}[ht!]
%\begin{center}
%\includegraphics[width=90mm]{fig3.eps}
%\caption{\footnotesize The dispersion curves for longitudinal modes for various temperatures above and below the critical temperature $T_c$ where Bose-Einstein condensation occurs, for $T>T_c$ (solid and dotted-dashed line), $T<T_c$ (dotted line) and $T=0$ (dashed line). The dotted-dashed line for $T>Tc$ is discarded. }
%\label{fig3}
%\end{center}
%\end{figure}
\begin{figure}[ht!]
	\begin{center}
		\includegraphics[width=60mm]{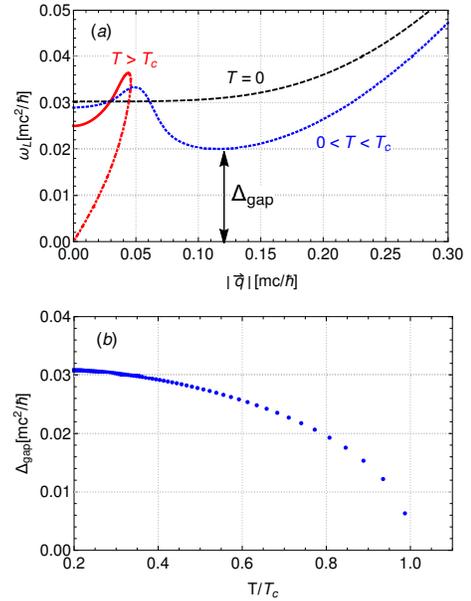}
		\caption{\footnotesize (a) Dispersion curves for longitudinal modes for $T>T_c$ (solid and dotted-dashed line), $0<T<T_c$ (dotted line), and $T=0$ (dashed line). The dotted-dashed line for $T>T_c$ is discarded; (b) Roton gap energy as a function of the relative temperature. Calculations were performed for $\eta=10^{-2} (mc/\hslash)^3$, and $T_c=0.14 mc^2/k_b$. }
		\label{fig4}
	\end{center}
\end{figure}

In the present study of a {\sl charged} relativistic Bose gas, the dispersion relation of the longitudinal plasmon mode shows an {\it ordinary} plasmon, near $|\vec{q}|=0$, and a {\it roton} excitation, near the value of $|\vec{q}|$ that corresponds to the local minimum. As the temperature is increased, the gap energy of the local minimum decreases, and Fig.\ref{fig4}(b) shows how it depends on the temperature. We may view the gap energy as an order parameter, which vanishes at $T=T_c$. For $0<T<T_c$, as the temperature is increased, the {\sl ordinary} plasmon near $|\vec{q}|=0$ turns into an elementary {\sl roton} excitation with a (lower) minimum energy at finite momentum. We further illustrate the presence of rotons by presenting the dispersion curves of the longitudinal mode for various densities at temperatures below the critical temperature $T_c$ in Fig.\ref{fig5}, for non-relativistic densities [Fig.\ref{fig5}(a) and (b)], and for relativistic and ultra-relativistic densities [Fig.\ref{fig5}(c) and (d)]. It should be noted that those results agree remarkably well with the non-relativistic results reported in the literature, obtained via a different route \cite{tosi}. 

Figures \ref{fig4} and \ref{fig5} suggest that thermal effects induce the rotons that will contribute to disorder the system. Thermally induced rotons were also present in the bosonic spectrum of a two-dimensional dilute Bose gas \cite{nogueira2006}, where it was argued that their emergence is a consequence of the strong phase fluctuation in two dimensions.
\begin{figure}[ht!]
	\begin{center}
		\includegraphics[width=80mm]{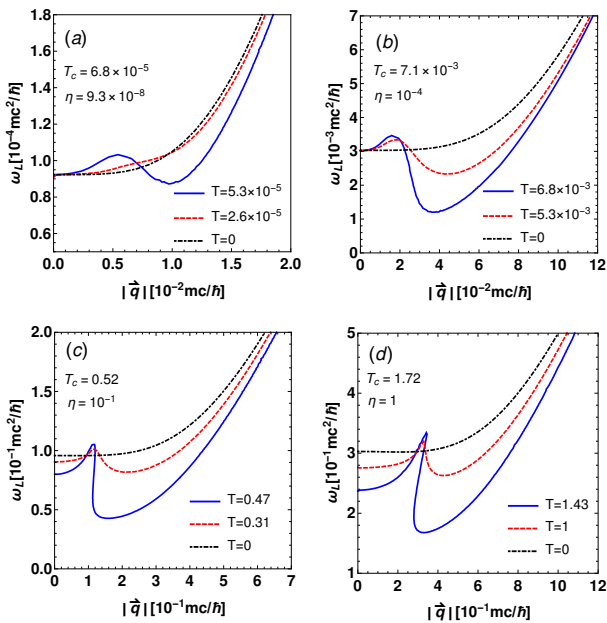}
		\caption{\footnotesize Dispersion curves for longitudinal modes for various temperatures below the critical temperature $T_c$ (in units of $mc^2/k_B$), and  various densities $\eta$ [in units of $(mc/\hslash)^3$]. In (a) and (b), calculations were performed for non-relativistic densities and temperatures; in (c) and (d), for relativistic and ultra-relativistic.}
		\label{fig5}
	\end{center}
\end{figure}

We remark that the Lagrangian density in our approach does not include a self-interaction term $\lambda(\phi^*\phi)^2$. The condensate is introduced by performing the substitution (\ref{nB1}), which yielded $T=0$ results for the longitudinal plasmon mode in agreement with the literature in both relativistic \cite{kowalenko1985response} and nonrelativistic limits \cite{hore1975}. We plan to include self-interactions, either directly or induced by integration over quantum electromagnetic fields, in a future investigation. In the non-relativistic limit, that inclusion will be useful in the description of the superfluidity of liquid Helium at low temperatures \cite{schmitt2015introduction}. In the relativistic limit, it could possibly help in the study of the relativistic Bose plasma found in astrophysical scenarios such as neutron stars, where the creation of charged pion pairs and pion condensation may take place \cite{kowalenko1985response,Moitra}. 
%the superfluid phase emerging from the breaking of $U(1)$ symmetry and formation of a Bose-Einstein %condensate (BEC), responsible for frictionless flow \cite{alford2013complex}.

\acknowledgments
The authors would like to thank the Brazilian Agencies CNPq and CAPES for partial financial support.


\begin{thebibliography}{99}


\bibitem{lieb} M. Grether, M. de Lanno, and G. A. Baker, Physical Review Letters \textbf{99}, 200406 (2007); A. Haber and H. Weldon, Physical Review Letters \textbf{46}, 1947 (1981); C. A. A. de Carvalho and S.G. Rosa, Journal of Physics A \textbf{13}, 3233 (1980).

%\bibitem{schafroth1955superconductivity} M. R.	Schafroth, Physical Review \textbf{100}, 463 (1955).

%\bibitem{alford2013complex} Alford, M. G., Mallavarapu, S. K., Schmitt, A. and Stetina, S., {\it Physical Review D}, \textbf{87}, 065001 (2013).

%\bibitem{baym1975neutron} G. Baym and P. Christopher, Annual Review of Nuclear Science \textbf{25}, 27-77 (1975).

%\bibitem{ginzburg1969superfluidity} V. L. Ginzburg,  Journal of Statistical Physics \textbf{1}, 3-24 (1969). 

\bibitem{AragaoPRDS2016} C. A. A. de Carvalho, Physical Review D \textbf{93}, 105005 (2016).

\bibitem{ReyezEPL2016} E. Reyes-G\'omez, L. E. Oliveira, and C. A. A. de Carvalho, Europhysics Letters \textbf{114}, 17009 (2016).

\bibitem{JPP} C. A. A. de Carvalho and D. M. Reis, Journal of Plasma Physics \textbf{84}, 905840112 (2018).

\bibitem{ReisAdP2018} D. M. Reis, E. Reyes-G\'omez, L. E. Oliveira, and 
C. A. A. de Carvalho, Annalen der Physik \textbf{530}, 1700443 (2018).

\bibitem{AragaoJOSA2020} C. A. A. de Carvalho, D. M. Reis, and D. Szilard,  \href{http://arxiv.org/abs/2006.03129}{arXiv:2006.03129} (2020).

\bibitem{Veselago} V. G. Veselago, Soviet Physics Uspekhi \textbf{10}, 509 (1968).

\bibitem{Pendry} J. B. Pendry, A. J. Holden, W. J. Steward, and I. Youngs, Physical
Review Letters \textbf{76}, 4773 (1996); D. R. Smith, W. J. Padilla, D. C. Vier, S. C. Nemat-Nasser, and
S. Schultz, Physical Review Letters \textbf{84}, 4184 (2000).  

\bibitem{engheta} N. Engheta and R.W. Ziolkowski, \textit{Metamaterials: Physics and Engineering Explorations}, IEEE Wiley, New York (2006).

\bibitem{handbook} {\it Metamaterials Handbook: 
Vol. I. Phenomena and Theory of Metamaterials}, ed. F. Capolino, Taylor and Francis, Boca Raton (2009).

\bibitem{clarendon} D. G. Henshaw and A. D. B. Woods, Physical Review 
\textbf{121}, 1266 (1961); H. R. Glyde, in \textit{Excitations 
in Liquid and Solid Helium}, Clarendon (1994).

\bibitem{landau} L. D. Landau, USSR Journal of Physics \textbf{11}, 91–92 (1947).

\bibitem{Kapusta} J. I. Kapusta and C. Gale, {\it Finite-Temperature Field Theory}, Cambridge University Press, Cambridge, England (2006).

\bibitem{AP} I. A. Akhiezer  and S. V. Peletminskii, Soviet Physics JETP \textbf{11}, 1316 (1969); E. S. Fradkin, {\it Proceedings of the P. N. Lebedev Physics Institute}, vol. 29, Consultants Bureau (1967).

\bibitem{IZ}  C. Itzykson and J. B. Zuber, {\it Quantum Field Theory}, McGraw-Hill, New-York (1980).

\bibitem{melrose} D. Melrose, {\it Quantum plasmadynamics: magnetized plasmas}, Springer (2012).

\bibitem{hore1975} S. R. Hore and  N. E. Frankel, Physical Review B \textbf{12}, 2619 (1975).



\bibitem{santos} L. Santos, G. V. Shlyapnikov, and M. Lewenstein, 
Physical Review Letters \textbf{90}, 250403 (2003).

\bibitem{fischer} U. R. Fischer, Physical Review A \textbf{73}, 031602(R)(2006).

\bibitem{wilson} R. M. Wilson, S. Ronen, J.L. Bohn, and H. Pu, Physical Review Letters \textbf{100}, 245302 (2008).

\bibitem{lahaye} T. Lahaye, C. Menotti, L. Santos, M. Lewenstein, and T. Pfau, 
Reports of Progress in Physics \textbf{72}, 126401 (2009).

\bibitem{bisset} R.N. Bisset and P.B. Blakie, Physical Review Letters \textbf{110}, 265302 (2013).

\bibitem{odell} D.H.J. O’Dell, S. Giovanazzi, and G.Kurizki, Physical Review Letters \textbf{90}, 110402 (2003). 

\bibitem{henkel} N. Henkel, R. Nath, and T. Pohl,  Physical Review Letters \textbf{104}, 195302 (2010).

\bibitem{chomaz2018observation} L. Chomaz, R. M. van Bijnen, 
D. Petter, G. Faraoni, S. Baier, J.H. Becher, and F. Ferlaino,
Nature Physics \textbf{14}, 442 (2018); D. Petter, G. Natale, 
R.M.W. van Bijnen, A. Patscheider, M. J. Mark, L. Chomaz, and 
F. Ferlaino, Physical Review Letters \textbf{122}, 183401 (2019).

%\bibitem{Pines} D. Pines and P. Nozi\`eres, P., \textit{The Theory 
%of Quantum Liquids}, Benjamin (1966); D. J. Thouless, \textit{The Quantum Mechanics of Many-body %Systems}, Academic Press (1972); Glyde, H. R., \textit{Excitations in Liquid and Solid 
% Helium}, Clarendon (1994).
 
 

\bibitem{hirashima1985} Hirashima, S. Dai, and H. Namaizawa, Progress of Theoretical Physics \textbf{74}, 400-404(1985).

\bibitem{zhang1995} S. C. Zhang, {\it Low-Dimensional Quantum Field Theories for Condensed Matter Physicists}, p. 191-224 (1995).

\bibitem{tosi} B. Davoudi and M. P. Tosi, Phys. Rev. B \textbf{72}, 134520 (2005).

\bibitem{nogueira2006} F. S. Nogueira and H. Kleinert, Physical Review B \textbf{73}, 104515 (2006).


\bibitem{kowalenko1985response} V. Kowalenko, N. E. Frankel, and K. C. Hines, Physics Reports \textbf{126}, 109-187 (1985).

\bibitem{schmitt2015introduction} A. Schmitt, Lecture Notes in Physics \textbf{888}, Springer (2015).

\bibitem{Moitra} P. Moitra, Y. Yang, Z. Anderson, I. I. Kravchenko, D. P. Briggs, and J. Valentine, 
Nature Photonics \textbf{7}, 791 (2013).




%\bibitem{ramakrishna} S. A. Ramakrishna, Rep. Prog. Phys. \textbf{68}, 449 (2005).

%\bibitem{wegener} M. Kadic, T. Bückmann, R. Schittny, and M. Wegener, 
%Rep. Prog. Phys. \textbf{76}, 126501 (2013). 


%\bibitem{han2012electron} W.B. Han, R. Ruffini and S.S. Xue,  Phys. Rev. D, \textbf{86}, 084004 (2012).

%\bibitem{berezhiani2015}  V.I. Berezhiani, N.L. Shatashvili and N.L. Tsintsadze, Phys. Scr. \textbf{90}, 068005 (2015).

%\bibitem{Nerush2011} E. N. Nerush, I. Y. Kostyukov, A. M. Fedotov, N. B. Narozhny, N. V. Elkina, and H. Ruhl,  Phys. Rev. Lett. \textbf{106}, 035001 (2011).

%\bibitem{raicher2014}  E. Raicher, S. Eliezer and A. Zigler, Phys. of Plasmas \textbf{21}, 053103 (2014).

%\bibitem{raicher2016} E. Raicher, S. Eliezer and A. Zigler. Phys. Rev. A \textbf{94}, 062105 (2016).

%\bibitem{Aamodt2010} K. Aamodt \textit{et al}. (The Alice Collaboration) Phys. Rev. Lett. \textbf{105}, 252302 (2010); Phys. Rev. Lett. \textbf{105}, 252301 (2010); Phys. Rev. Lett. \textbf{106}, 032301 (2011). 


%\bibitem{Yousuf2017} J. M. Yousuf, S. Mitra and V. Chandra,  Phys. Rev. D \textbf{95}, 094022 (2017)
 

%\bibitem{Smith}  D. R. Smith, Willie J. Padilla, D. C. Vier, S. C. Nemat-Nasser, and S. Schultz, Phys. Rev. Lett. \textbf{84}, 4184 (2000).

%\bibitem{Zhe} N. I. Zheludev, Science \textbf{348}, 973 (2015) and references therein.

%\bibitem{Lindhard1954} J. Lindhard, K. Dan. Vidensk. Selsk., Mat. Fys. Medd. \textbf{28}, 8 (1954).

%\bibitem{Jensen} E. T. Jensen, R. E. Palmer, W. Allison, and J. F. Annett, Phys. Rev. Lett. \textbf{66}, 492 (1991).

%\bibitem{Portail} M. Portail, M. Carrere, and J. M. Layet, Surf. Sci. \textbf{433-435}, 863 (1999).

%\bibitem{Zou} X. Zou, J. Luo, D. Lee, C. Cheng, D. Springer, S. K. Nair, S. A. Cheong, H. J. Fan, and E. E. M. Chia, J. Phys. D: Appl. Phys. \textbf{45},  465101 (2012).





%\bibitem{Bohm} D. Bohm and E. P. Gross, Phys. Rev. \textbf{75}, 1851 (1949); {\it ibid}., Phys. Rev. \textbf{75}, 1864 (1949).



%\bibitem{landau1941} Landau, L. {\it Physical Review}, v. 60, n. 4, p. 356, 1941.
	%\bibitem{Smith} Smith, D. R., Padilla, W. J., Vier, D. C., Nemat-Nasser, S. C. and Schultz, S. {\it Physical Review Letters}, v. 84, n. 18, p. 4184, 2000.

%\bibitem{Zhe} Zheludev, N. I. {\it Science}, v. 348, n. 6238, p. 973-974, 2015.

%\bibitem{yan2015hydrodynamic} W. Yan, Phys. Rev. B, \textbf{91}, 11, 115416 (2015).
%\bibitem{govorov2014photogeneration} A. O. Govorov, H. Zhang, H.V. Demir and Y.K. Gun'ko.  Nano Today \textbf{9}, 85â101 (2014).


%\bibitem{AP} I. A. Akhiezer  and S. V. Peletminskii, {\it Sov. Phys. JETP}, v. 11, n.6, p. 1316, 1969; E. S. Fradkin, {\it Proceedings of the P. N. Lebedev Physics Institute}, vol. 29, Consultants Bureau, 1967.

%\bibitem{jancovici1962relativistic}	Jancovici, B. {\it Il Nuovo Cimento (1955-1965)}, v. 25, n. 2, p. 428-455, 1962.

%\bibitem{melrose1984dispersion}	Melrose, D. B., and  Hayes, L. M. {\it  Australian Journal of Physics}, v. 37, n. 6, p. 639-650, 1984.

%\bibitem{pulsifer1992pair} Pulsifer, P., and G. Kalman. {\it Physical Review A}, v. 45, n.8, p, 5820, 1992. 


%\bibitem{Zou} X. Zou, J. Luo, D. Lee, C. Cheng, D. Springer, S. K. Nair, S. A. Cheong, H. J. Fan, and E. E. M. Chia, {\it J. Phys. D: Appl. Phys}, v. 45, n. 46,  p. 465101, 2012.

 




 



\end{thebibliography}
\end{document}